\documentclass[conference, 10pt]{IEEEtran}
\IEEEoverridecommandlockouts
\usepackage{cite}
\usepackage{amsmath,amssymb,amsfonts, bbold}
\usepackage{mathtools}
\usepackage{algorithm}
\usepackage{algorithmic}
\usepackage[font=footnotesize]{caption}
\usepackage[font=footnotesize]{subcaption}
\usepackage{graphicx}
\usepackage{textcomp}
\usepackage{xcolor}
\usepackage{xfrac}
\usepackage{booktabs}
\usepackage{comment}

\usepackage{pgfplots}
\pgfplotsset{compat=1.16}
\usepgfplotslibrary{fillbetween}
\pgfplotsset{yticklabel style={text width=1.2em,align=right}}
\definecolor{matblue}{rgb}{0, 0.4470, 0.7410}
\definecolor{matred}{rgb}{0.85, 0.3250, 0.0980}
\definecolor{matorange}{rgb}{0.9290, 0.6940, 0.1250}
\definecolor{matviolet}{rgb}{0.4940, 0.1840, 0.5560}
\definecolor{matgreen}{rgb}{0.4660, 0.6740, 0.1880}
\definecolor{matskyblue}{rgb}{0.3010, 0.7450, 0.9330}

\newcommand{\vect}[1]{\boldsymbol{ #1 }}

\newtheorem{theorem}{Theorem}

\begin{document}

\title{Stochastic Resource Allocation via\\ \hspace{3.5pt} Dual Tail Waterfilling
\thanks{This work is supported by the NSF under grant CCF 2242215.}
}

\author{\IEEEauthorblockN{Gokberk Yaylali and Dionysis Kalogerias}
\IEEEauthorblockA{\textit{Department of Electrical Engineering} \\
\textit{Yale University}}
}

\maketitle

\begin{abstract}

Optimal resource allocation in wireless systems still stands as a rather challenging task due to the inherent statistical characteristics of channel fading. On the one hand, minimax/outage-optimal policies are often overconservative and analytically intractable, despite advertising maximally reliable system performance. On the other hand, ergodic-optimal resource allocation policies are often susceptible to the statistical dispersion of heavy-tailed fading channels, leading to relatively frequent drastic performance drops. We investigate a new \textit{risk-aware} formulation of the classical stochastic resource allocation problem for point-to-point power-constrained communication networks over fading channels with no cross-interference, by leveraging the Conditional Value-at-Risk (CV@R) as a coherent measure of risk. We rigorously derive closed-form expressions for the CV@R-optimal risk-aware resource allocation policy, as well as the optimal associated quantiles of the corresponding user rate functions by capitalizing on the underlying fading distribution, parameterized by dual variables. We then develop a purely dual tail waterfilling scheme, achieving significantly more rapid and assured convergence of dual variables, as compared with the primal-dual tail waterfilling algorithm, recently proposed in the literature. The effectiveness of the proposed scheme is also readily confirmed via detailed numerical simulations.
\end{abstract}
\vspace{2pt}
\begin{IEEEkeywords}
Resource Allocation, Conditional Value-at-Risk (CV@R), Waterfilling, Risk-Aware Optimization, Dual Descent.
\end{IEEEkeywords}

\vspace{-2pt}
\section{Introduction}
\vspace{-2pt}
% Insert RA introduction here 
In this paper, we revisit the classical resource allocation problem in point-to-point communication networks with no cross-interference operating over random fading channel realizations $\vect{H} \in \mathcal{H} \subseteq \mathbb{R}^{N_U}$. In the dynamic landscape of wireless networks, efficiently allocating resources stands as a critical and perpetual challenge to ensure optimal and robust system performance. In fact, even achieving decent performance in expectation is often insufficient in modern networking applications, as occurrence of less-probable though statistically significant fading events might prompt rather unsatisfactory outcomes \cite{tsiamis2020risk}. To this extent, heavy-tailed characteristics of channel fading necessitate the development of statistically robust resource allocation policies to compensate such non-typical events, even at the cost of minor performance degradation on average.

Conventionally, allocation of resources, such as transmission power and/or channel access, is carried out by either deterministic or stochastic methods to optimize certain network utilities. In the deterministic framework, including most conservative minimax formulations \cite{Parsaeefard2013, Mokari2016}, the statistical variability of fading is often disregarded as an essential characteristic of the system. On the other hand, stochastic approaches consider expectations of random network objectives \cite{Ribeiro2012, Mokari2016, AliHemmati2018, Kalogerias2020, Hashmi2021} (e.g., transmission rates) while aiming to maximally optimize performance in the long-term, i.e., in the ergodic sense.

% Insert weak aspects of current methods here
While minimax-type resource allocation policies are often regarded as ``robust'' due to their maximally reliable system performance \cite{Parsaeefard2013, Mokari2016}, they are, in fact, overcautious and exhibit conservative system performance. Such policies target the ``worst-case scenario'', inherently preventing the system to achieve higher average network utilities, e.g., transmission rates. On the other extent, ergodic resource allocation policies are optimal only in expectation, and generally fail to effectively anticipate comparably rare-occurring but operationally significant channel observations, e.g., deep fades. Such fading realizations are rather observable in communication media with heavy-tailed fading distributions, leading to severe service outages. In fact, it is well-known that ergodic policies are typically channel-opportunistic \cite{Ribeiro2012}, subsequently leading to poor performance over sporadic channel realizations. This corresponds to considerable operational spectrum underutilization, correlating with unreliable system performance.

Although approaches based on outage probability optimization \cite{Li2005mac} aim to overcome the issues presented by the methods above, they ultimately raise new questions: How do we select feasible outage probability targets to effectively allocate resources, and even when those targets are feasible, how do we guarantee that they prompt substantial system performance? Quantile-based resource allocation policies, including outage rate/capacity optimization, aim to alleviate those questions, however, they are limited in terms of interpretability, and inherently lack favorable structure, such as convexity. 

% Insert introduction to risk-aware approaches, extension & contributions here
Risk-aware approaches are steadily becoming important \cite{Vu2018, Li2021, Bennis2018, Kalogerias2022cvar}, particularly in modern network applications necessitating strict reliability requirements to be met. To this end, we investigate a risk-aware problem formulation of the resource allocation problem in multi-terminal point-to-point resource-constrained communication network with no cross-interference by utilizing the Conditional-Value-at-Risk (CV@R) as a measure of fading risk \cite{Rockafellar2000}. CV@R is a coherent risk measure \cite{Shapiro2014} continuously spanning between the extremes of ergodic and minimax settings, allowing us to reformulate the resource allocation problem as a convex, interpretable, and well-structured extension of its classical (ergodic) counterpart \cite{Cover2005, Ribeiro2012}. In our previous work \cite{yaylali2023cvar}, we introduced the \textit{primal-dual tail waterfilling (PDTW) algorithm} for purely data-driven CV@R-optimal risk-aware resource allocation policy learning, achieving fully tunable system robustness and reliability. 

In this paper, we exploit potential prior information on the fading probability distribution (available --even approximately-- in various settings), leading to the \textit{dual tail waterfilling (DTW) algorithm}, facilitating globally optimal, statistically robust and reliable risk-aware resource policy optimization. Our contributions are as follows: We rigorously obtain closed-form expressions of the CV@R-optimal Lagrangian-relaxed risk-aware resource policy, as well as the related quantile measures for user rates regulated by CV@R-optimal resource policies (and the optimal rate vector). Then, we design a purely dual descent scheme (DTW) to attain a globally optimal risk-aware policy in a recursive, subgradient-based fashion. Efficiently exploiting fading priors drastically accelerates convergence speed, as well as the overall effectiveness of our CV@R-based approach. We conduct detailed numerical simulations substantiating the effectiveness and good empirical characteristics of DTW algorithm for two common network utilities.

\section{Problem Formulation}

We consider a $N_U$-terminal parallel point-to-point communication channel model with no cross-interference. Also, for simplicity, we assume perfect channel state information (CSI) at transmission time. The resources are allocated via a policy function $\vect{p}(\vect{h}) \succeq \vect{0}$, where $\vect{h}$ is the instantaneous fading vector, whose elements $h_i,\ i \in \{1, \dots, N_U\}$ correspond to fading coefficients of parallel links, distributed by a cumulative distribution function (cdf) $F_{h_i}$. The instantaneous transmission rate of communication link $i \in \{1, \dots, N_U\}$ in the network is
\begin{equation}
\label{eqn:rate_function}
r_i(p_i(h_i), h_i) \triangleq \log \left( 1 + \frac{p_i(h_i) \cdot h_i^2}{\sigma_i^2} \right),
\end{equation}
where $\sigma_i^2$ is the noise variance of the corresponding link. In an ergodic setting, optimal resource policies can be readily obtained by solving a classical stochastic problem \cite{Ribeiro2012, Cover2005}.
%\cite{Ribeiro2012} by solving the classical problem of
%\begin{equation}
%\begin{aligned}
%P^* = \underset{\vect{x} \in \mathcal{X}, \vect{p}\succeq \vect{0}}{\mathrm{maximize}} \quad & f_0 (\vect{x}),\\
%\mathrm{subject\ to} \quad & \vect{x} \preceq \mathbb{E}\left[ \vect{r}(\vect{p}(\vect{h}),\vect{h}) \right],\\
%& \Vert \mathbb{E}\left[ \vect{p}(\vect{h}) \right] \Vert_1 \leq P_0,
%\end{aligned}
%\end{equation}
%where $f_0$ is a given concave objective function, $\vect{x}$ is mean-ergodic rate vector, $\mathcal{X}$ is a convex set, $\vect{r}$ is the instantaneous rate vector, $P_0$ is the total mean power budget. 
To meaningfully mitigate the adverse effects of commonly dispersive or heavy-tailed channel fading in system performance, we investigate a risk-aware extension of the resource allocation problem formulated as \cite{yaylali2023cvar} 
\begin{equation}
\label{eqn:prob_formulation_cvar}
\begin{aligned}
P^* = \underset{\vect{x} \in \mathcal{X}, \vect{p}\succeq \vect{0}}{\mathrm{maximize}} \quad & f_0 (\vect{x}),\\
\mathrm{subject\ to} \quad & \vect{x} \preceq -\text{CV@R}^{\vect{\alpha}}\left[ -\vect{r}(\vect{p}(\vect{h}),\vect{h}) \right],\\
& \Vert \mathbb{E}\left[ \vect{p}(\vect{h}) \right] \Vert_1 \leq P_0,
\end{aligned}
\end{equation}
where $\vect{x}$ is a \textit{risk-ergodic} rate vector, and CV@R is defined as
\begin{equation}
\label{eqn:cvar_inf}
\text{CV@R}^{\alpha}[z] \triangleq \inf_{t \in \mathbb{R}}\ t + \dfrac{1}{\alpha} \mathbb{E}\left[ (z - t)_+ \right],
\end{equation}
for an integrable random cost $z$, $\alpha \in (0,1]$ is the corresponding confidence level, the vector notation $\text{CV@R}^{\vect{\alpha}}[\cdot]$ represents elementwise operations (with $\vect{\alpha}$ being a vector of corresponding CV@R confidence levels), and $(\cdot)_+ = \max\{\cdot, 0\}$. Note that CV@R is a convex, monotone, translation equivariant and positively homogeneous --therefore coherent-- \textit{risk measure} \cite{Shapiro2014}, strictly generalizing expectation in a tunable and tractable fashion, since it satisfies
\begin{equation}
\begin{aligned}
\text{CV@R}^0 [z] &= \lim_{\alpha \to 0}\ \text{CV@R}^{\alpha} [z] = \mathrm{ess}\ \sup z,\\
\text{CV@R}^1 [z] &= \mathbb{E}\left[ z \right] \leq \text{CV@R}^\alpha [z] \text{ for } \alpha \in (0,1],
\end{aligned}
\end{equation}
also being monotonic in $\alpha$. CV@R measures the expected loss of a random cost $z$ \textit{restricted} to the upper tail of the underlying distribution, of probability equal to $\alpha$ \cite{Rockafellar2000}. We modify the formulation in \eqref{eqn:cvar_inf} to measure an expected \textit{reward} constrained in the \textit{lower} tail of probability equal to $\alpha$, suitable for maximizing objectives, as
\begin{equation}
\label{eqn:cvar_sup}
-\text{CV@R}^{\alpha}[-z] \triangleq \sup_{t \in \mathbb{R}}\ t - \dfrac{1}{\alpha} \mathbb{E}\left[ (t - z)_+ \right].
\end{equation}
Utilizing \eqref{eqn:cvar_sup} in \eqref{eqn:prob_formulation_cvar}, we may simply express the risk-aware resource allocation problem as
\begin{equation}
\label{eqn:prob_formulation_cvar_explicit}
\begin{aligned}
P^* = \underset{\vect{x} \in \mathcal{X}, \vect{p}\succeq \vect{0}, \vect{t} }{\mathrm{maximize}} \quad & f_0 (\vect{x}),\\
\mathrm{subject\ to} \quad & \vect{x} \preceq \vect{t} - \dfrac{1}{\vect{\alpha}} \odot \mathbb{E}\left[ (\vect{t} - \vect{r}(\vect{p}(\vect{h}), \vect{h}) )_+ \right],\\
& \Vert \mathbb{E}\left[ \vect{p}(\vect{h}) \right] \Vert_1 \leq P_0,
\end{aligned}
\end{equation}
where ``$\odot$'' stands for Hadamard product, and division with respect to vector $\vect{\alpha}$ similarly stands for elementwise division. %, where V@R is defined as
%\begin{equation}
%\label{eqn:var_uppertail}
%\text{V@R}^{\alpha}[z] \triangleq \inf\left\{ t\ \vert\ \mathbb{P}\{ z > t \} \leq \alpha \right\},
%\end{equation}
%for an integrable random cost $z$. Notice that V@R \eqref{eqn:var_uppertail} measures the upper tail $\alpha$-quantile of the corresponding distribution. Similar to CV@R, we are concerned with $-\text{V@R}[-z]$ formulation for the lower tail $\alpha$-quantile. Notice that the optimal $\vect{t}^*$ corresponds to the instantaneous rate V@R. 
Problem \eqref{eqn:prob_formulation_cvar_explicit} remains convex due to the inherent coherence of CV@R. Nonetheless, problem \eqref{eqn:prob_formulation_cvar_explicit} is still rather complicated, since the fading vector $\vect{h}$ attains values of a continuum, introducing infinite-dimensionality to the problem, therefore solving \eqref{eqn:prob_formulation_cvar_explicit} may seem an obscure and difficult challenge. However, under the assumption of certain constraint qualifications, such as Slater's condition, strong Lagrangian duality in \eqref{eqn:prob_formulation_cvar_explicit} is observed --hence, there is no duality gap. This fact enables the use of the dual problem of \eqref{eqn:prob_formulation_cvar_explicit} within the Lagrangian duality framework. The Lagrangian of \eqref{eqn:prob_formulation_cvar_explicit} is defined as
\begin{equation}
\begin{multlined}
\mathcal{L}(\vect{x}, \vect{p}, \vect{t}, \vect{\Lambda}) \triangleq f_0(\vect{x}) + \mu \left( P_0 - \Vert \mathbb{E}\left[ \vect{p}(\vect{h}) \right] \Vert_1 \right)\\
+ \vect{\lambda}^T \left( \vect{t} - \frac{1}{\vect{\alpha}} \odot \mathbb{E}\left[ (\vect{t} - \vect{r}(\vect{p}(\vect{h}), \vect{h}) )_+ \right] - \vect{x} \right),
\end{multlined}
\end{equation}
where $\vect{\Lambda} = \left( \vect{\lambda}, \mu \right) \succeq \vect{0}$ are the Lagrangian coefficients --dual variables-- for corresponding constraints in \eqref{eqn:prob_formulation_cvar_explicit}. The dual function is then expressed as the maximization of the Lagrangian function over the primal variable triplet $(\vect{x}, \vect{p}, \vect{t})$, i.e.,
\begin{equation}
q(\vect{\Lambda}) \triangleq \sup_{\vect{x} \in \mathcal{X}, \vect{p}\succeq \vect{0}, \vect{t}}\ \mathcal{L}(\vect{x}, \vect{p}, \vect{t}, \vect{\Lambda}).
\end{equation}
We may subsequently define the dual problem as the minimization of the dual problem with respect to dual variables, i.e.,
\begin{equation}
\label{eqn:dual_prob_formulation}
\begin{aligned}
%D^* &= \inf_{\vect{\Lambda} \succeq \vect{0}}\ q(\vect{\Lambda}),\\
D^* &\triangleq \inf_{\vect{\Lambda} \succeq \vect{0}}\ \sup_{\vect{x} \in \mathcal{X}, \vect{p}\succeq \vect{0}, \vect{t}}\ \mathcal{L}(\vect{x}, \vect{p}, \vect{t}, \vect{\Lambda}).
\end{aligned}
\end{equation}
%The weak duality between the primal and dual problems axiomatically invokes $P^* \leq D^*$. Under the assumption of Slater's condition, the primal problem \eqref{eqn:prob_formulation_cvar_explicit} and the dual problem \eqref{eqn:dual_prob_formulation} exhibit no duality gap --strong duality is observed, thus $P^* = D^*$. 
Recall that the primal problem \eqref{eqn:prob_formulation_cvar_explicit} exhibits no duality gap and remains infinite-dimensional, however, the corresponding dual problem is finite-dimensional, initiating the use of \eqref{eqn:dual_prob_formulation} for globally optimal solutions as anticipated. We next propose an efficient dual waterfilling scheme (cf. PDTW algorithm of \cite{yaylali2023cvar}) to solve the minimax problem in \eqref{eqn:dual_prob_formulation}, and obtain dual variable-parameterized closed-form solutions of \textit{all} primal variables involved, including the CV@R-optimal solution to the risk-aware resource policy.

\section{The Dual Tail Waterfilling}

The dual problem \eqref{eqn:dual_prob_formulation} can be separated into several subproblems with respect to the primal variables. Leveraging the interchangeability principle \cite[Theorem 7.92]{Shapiro2014}, we may express the dual problem in the form
\vspace{-6pt}
\begin{multline}
\label{eqn:dual_prob_explicit}
\hspace{-10bp} D^* = \inf_{\vect{\Lambda} \succeq \vect{0}}\ \mu P_0 + \sup_{\vect{x} \in \mathcal{X}} \left\{ f_0(\vect{x}) - \vect{\lambda}^T \vect{x} \right\} +\sup_{\vect{t} \in \mathbb{R}^{N_U}} \Bigg\{ \sum_{i=1}^{N_U} \lambda_i t_i \\
\hspace{-3bp}+ \mathbb{E}\left[ \sup_{p_i \geq 0} \left\{ - \left( \frac{\lambda_i}{\alpha_i} (t_i -r_i(p_i,h_i) )_+ \right) - \mu p_i \right\} \right] \Bigg\}.\hspace{-3bp}
\end{multline}
Next, by capitalizing on the separation of subproblems, we rigorously derive the closed-form solution of all primal variables, particularly the dual variable-parameterized CV@R-optimal resource policy and the corresponding optimal $\vect{t}^*$.

\subsection{CV@R-Optimal Risk-Aware Resource Policy}

The particular resource policy subproblem for each terminal $i \in \{1, N_U \}$ is
\begin{equation}
\label{eqn:p_subproblem}
\sup_{p_i \geq 0} \left\{ - \left( \frac{\lambda_i}{\alpha_i} (t_i -r_i(p_i,h_i) )_+ \right) - \mu p_i \right\}.
\end{equation}
Next, we present the optimal solution to \eqref{eqn:p_subproblem}, exhibiting the unique behavior of optimal risk-aware policy, compared to its risk-neutral (classical) counterpart.
\begin{theorem}[CV@R-Optimal Risk-aware Policy]
\label{thr:optimal_p_cvar}
An optimal solution to the resource policy subproblem \eqref{eqn:p_subproblem} for terminal $i \in \{1, \dots, N_U\}$ is
\begin{equation}
\label{eqn:p_closed_form}
\boxed{p_i^*(h_i,\cdot) \triangleq \min\left\{ \left( \frac{\lambda_i}{\mu \alpha_i} - \frac{\sigma_i^2}{h_i^2} \right)_+,\ \frac{\sigma_i^2 \left(e^{(t_i)_+} - 1 \right)}{h_i^2} \right\},}
\end{equation}
whenever $(\lambda_i, \mu) \neq \vect{0}$, otherwise selecting $p_i^* = 0$ is optimal.
\end{theorem}
\begin{IEEEproof}[Proof of Theorem \ref{thr:optimal_p_cvar}]
Notice that problem \eqref{eqn:p_subproblem} is concave, and becomes null when $(\lambda_i, \mu) = \vect{0}$. For $\lambda_i = 0$ or $t_i \leq 0$, and $\mu > 0$, the subproblem stands trivial with the optimal solution of $p_i^* = 0$. For $\lambda_i > 0$ and $\mu = 0$, the subproblem becomes
\begin{equation}
\sup_{p_i \geq 0}\ \left\{ - \frac{\lambda_i}{\alpha_i} \left( t_i- \log \left( 1 + \frac{p_i h_i^2}{\sigma_i^2} \right) \right)_+ \right\},
\end{equation}
and choosing $p^*_i = \sfrac{\sigma_i^2 \left(e^{(t_i)_+} - 1 \right)}{h_i^2}$ is optimal. For $(t_i, \lambda_i, \mu) \succ \vect{0}$ --assumed hereafter--, each subgradient $g(p_i)$ of the objective of \eqref{eqn:p_subproblem} can be expressed as
\begin{align}
\label{eqn:p_subgradient}
g(p_i,\cdot) \hspace{-1.5pt}=\hspace{-1.5pt} -\mu 
\hspace{-1pt}+\hspace{-1pt}
\frac{\lambda_i}{\alpha_i} H \left[ t_i
\hspace{-1pt}-\hspace{-1pt}
\log \hspace{-1pt} \left( 1 
\hspace{-1pt}+\hspace{-1pt}
\frac{p_i h_i^2}{\sigma_i^2} \right) \hspace{-1pt} \right] \frac{h_i^2}{\sigma_i^2 \hspace{-1pt}+\hspace{-1pt} p_i h_i^2},
\end{align}
where $H[\cdot]$ is any selection of the Heaviside step multifunction. Notice that $g$ is a decreasing function of $p_i \geq 0$, and the maximum value of subgradients is attained at $p_i = 0$, where
\begin{equation}
\Bar{g} = -\mu + \frac{\lambda_i}{\alpha_i} \frac{h_i^2}{\sigma_i^2},
\end{equation}
where $\Bar{g}$ is in the subdifferential of the objective of \eqref{eqn:p_subproblem} at $p_i = 0$. If $\Bar{g} \leq 0$, occurring iff $\frac{\lambda_i}{\mu \alpha_i} - \frac{\sigma_i^2}{h_i^2} \leq 0$, then the trivial choice $p_i^* = 0$ naturally becomes optimal. If $\Bar{g} > 0$, occurring iff $\frac{\lambda_i}{\mu \alpha_i} - \frac{\sigma_i^2}{h_i^2} > 0$, we exploit the fact that $0 \in \partial f(x^*)$ for an arbitrary function $f$ at the maximizing value $x^*$, and investigate two scenarios for a subgradient $g$ to attain zero. In the first scenario, suppose a $p^*_i \geq 0$ exists such that
\begin{equation}
\label{eqn:p_heaviside_1}
%\begin{aligned}
\textstyle H \left[ t_i- \log \left( 1 + \frac{p^*_i h_i^2}{\sigma_i^2} \right) \right] = 1 \Leftrightarrow \textstyle t_i - \log \left( 1 + \frac{p^*_i h_i^2}{\sigma_i^2} \right) > 0.
%\end{aligned}
\end{equation}
Then, from \eqref{eqn:p_subgradient}, we subsequently have
\begin{align}
\label{eqn:p_soln_first}
p^*_i(h_i,\cdot) &= \left( \frac{\lambda_i}{\mu \alpha_i} - \frac{\sigma_i^2}{h_i^2} \right)_+,
\end{align}
provided that $p_i^*$ satisfies \eqref{eqn:p_heaviside_1} as
\begin{align}
\label{eqn:p_branch_condition}
\frac{\sigma_i^2\left(e^{t_i} - 1 \right)}{h_i^2} > \left( \frac{\lambda_i}{\mu \alpha_i} - \frac{\sigma_i^2}{h_i^2} \right),
\end{align}
providing a \textit{branch condition}. For the second scenario, suppose a $p^*_i \geq 0$ exists such that
\begin{equation}
\label{eqn:p_heaviside_2}
%\begin{aligned}
\textstyle t_i - \log \left( 1 + \frac{p^*_i h_i^2}{\sigma_i^2} \right) = 0 \Leftrightarrow \textstyle H \left[ t_i- \log \left( 1 + \frac{p^*_i h_i^2}{\sigma_i^2} \right) \right] = C,
%\end{aligned}
\end{equation}
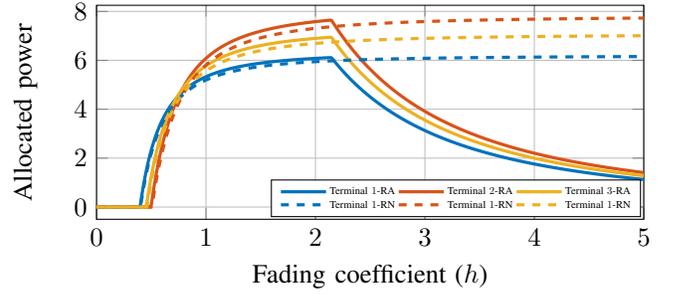
\begin{figure}[t]
\begin{tikzpicture}[trim axis right,baseline]
\begin{axis}[
    width=\linewidth,
    height=.5\linewidth,
    ylabel={Allocated power}, xlabel={Fading coefficient ($h$)},
    xmin=0, xmax=5,% xmajorticks=false,
    ymin=-0.5, ymax=8.25, ytick = {0, 2, 4, 6, 8},
    %legend image post style={scale=0.4},
    legend style={at={(0.995,0.01)},anchor=south east,
    nodes={scale=0.4, transform shape}},legend columns=3, grid]
\addplot[color=matblue, mark=no, very thick] table[x=fading,y=p_ra1] {Data/policy_function_of_h.txt};
\addplot[color=matred, mark=no, very thick] table[x=fading,y=p_ra2] {Data/policy_function_of_h.txt};
\addplot[color=matorange, mark=no, very thick] table[x=fading,y=p_ra3] {Data/policy_function_of_h.txt};
\addplot[color=matblue, mark=no, dashed, very thick] table[x=fading,y=p_rn1] {Data/policy_function_of_h.txt};
\addplot[color=matred, mark=no, dashed, very thick] table[x=fading,y=p_rn2] {Data/policy_function_of_h.txt};
\addplot[color=matorange, mark=no, dashed, very thick] table[x=fading,y=p_rn3] {Data/policy_function_of_h.txt};
\legend{Terminal $1$-RA, Terminal $2$-RA, Terminal $3$-RA, Terminal $1$-RN, Terminal $1$-RN, Terminal $1$-RN}
\end{axis}
\end{tikzpicture}
\caption{CV@R-Optimal resource allocation policies for risk-aware (RA, $\alpha = 0.90$) and risk-neutral (RN, $\alpha = 1.00$) settings in a $3$-terminal network.}
\label{policy_plot}
\end{figure}
where $C \in [0,1]$. Consequently, from \eqref{eqn:p_heaviside_2}, we have
\begin{equation}
\label{eqn:p_soln_second}
p^*_i(h_i,\cdot) = \frac{\sigma_i^2\left( e^{(t_i)_+} - 1 \right)}{h_i^2},
\end{equation}
provided that $p_i^*$ inherently satisfies
\begin{equation}
\label{eqn:p_comp_brach_condition}
\begin{aligned}
\frac{\sigma_i^2\left(e^{t_i} - 1 \right)}{h_i^2} &\leq \left( \frac{\lambda_i}{\mu \alpha_i} - \frac{\sigma_i^2}{h_i^2} \right)
\end{aligned}
\end{equation}
by combining \eqref{eqn:p_heaviside_2} and \eqref{eqn:p_subgradient}, meeting the complementary branch condition. Combining \eqref{eqn:p_soln_first}, \eqref{eqn:p_soln_second}, \eqref{eqn:p_branch_condition}, and \eqref{eqn:p_comp_brach_condition} ultimately concludes the proof.
\end{IEEEproof}

It follows that the optimal solution presented in \eqref{eqn:p_closed_form} is an extension of the risk-neutral resource allocation policy. Recall that CV@R is a tractable generalization of expectation at the extreme values of $\alpha$, i.e., $\alpha = 1$, leading $t_i$ to infinity. Therefore, the classical risk-neutral resource policy
\begin{equation}
{p_i}^N(h_i, \cdot) = \left( \frac{\lambda_i}{\mu} - \frac{\sigma_i^2}{h_i^2} \right)_+,
\end{equation}
stands within the operational spectrum of $\alpha$-parameterized risk-aware resource policy.

\subsection{Optimal Value-at-Risk / Risk-Ergodic Rate}

The remaining subproblems can be solved with respect to their corresponding primal variables. Recalling the dual problem \eqref{eqn:dual_prob_explicit}, we may now focus, for each terminal $i \in \{1, \dots, N_U\}$, on the subproblem
\begin{equation}
\label{eqn:t_subproblem}
%-\text{CV@R}^{\alpha_i}[-r_i] =
\hspace{-8pt}
\sup_{t_i \in \mathbb{R}} \hspace{-1pt} \left\{ G^* (t_i) \hspace{-1.5pt} \triangleq \hspace{-0.5pt} t_i \hspace{-1pt}-\hspace{-1pt} \frac{1}{\alpha_i} \mathbb{E}\left[ (t_i \hspace{-1pt}-\hspace{-1pt} r_i(p_i^*,h_i))_+ \right] \hspace{-1pt}-\hspace{-1pt}\frac{\mu}{\lambda_i}\mathbb{E}\left[ p_i^*\right] 
\right\}\hspace{-1pt},\hspace{-6pt}
\end{equation}
and we also define $G^*(\cdot)\triangleq G(p_i^*,\cdot)$. A closed-form expression for the optimal $t_i$ follows.
\begin{theorem}[Optimal Value-at-Risk]
\label{thr:optimal_var}
Let $F_{h_i}$ be a continuous and invertible cdf for the fading of terminal $i \in \{1, \dots, N_U\}$. Then, the optimal solution of \eqref{eqn:t_subproblem} at terminal $i$ is
\begin{equation}
\label{eqn:t_closed_form}
\begin{aligned}
\boxed{
t_i^*(\lambda_i,\mu) \triangleq \left( \log\left( \frac{\lambda_i}{\mu \alpha_i \sigma_i^2} \cdot \left( F_{h_i}^{-1}(\alpha_i) \right)^2 \right) \right)_+, }
\end{aligned}
\end{equation}
where $F_{h_i}^{-1}$ is the inverse of cdf $F_{h_i}$.
\end{theorem}
\begin{IEEEproof}[Proof of Theorem \ref{thr:optimal_var}]
Let $F_{r_i(t_i,\cdot)}$ be the cdf of instantaneous rate \eqref{eqn:rate_function} at terminal $i$, $i \in \{1, \dots, N_U\}$. Recall the optimal resource policy in \eqref{eqn:p_closed_form} to express $F_{r_i(t_i,\cdot)}$ as% the instantaneous rate as
%\begin{equation}
%\label{eqn:t_rate_as_func_of_h}
%\begin{aligned}
%r_i(t_i,\cdot) \triangleq \textstyle \min\left\{ \left( \log\left( \frac{\lambda_i}{\mu \alpha_i \sigma_i^2} \cdot h_i^2 \right) \right)_+ ,\ (t_i)_+ \right\}.
%\end{aligned}
%\end{equation}
%Employing \eqref{eqn:t_rate_as_func_of_h} to obtain $F_{r_i(t_i,\cdot)}$ gives
\begin{equation}
\label{eqn:t_cdf_R}
\begin{multlined}
F_{r_i(t_i,\cdot)}(r_i) \triangleq \mathbb{1}\left\{ (t_i)_+ < r_i \right\} \\
\textstyle +F_{h_i}\left( \sqrt{ \frac{\mu \alpha_i \sigma_i^2}{\lambda_i} \cdot e^{r_i} } \right) \cdot \mathbb{1}\left\{ 0 \leq r_i \leq (t_i)_+ \right\}.
\end{multlined}
\end{equation}
Note that $F_{r_i(t_i,\cdot)}$ also corresponds to the outage probability. %, which will be utilized in performance evaluation later on. 
Since $G$ in \eqref{eqn:t_subproblem} is \textit{jointly concave}, it can be shown that the subdifferential of $G^*$ may be characterized by
\begin{equation}
\label{eqn:t_subdiff_t}
\begin{aligned}
\partial_{t_i} \sup_{p_i \geq 0} G(p_i, t_i) &=  \partial_{t_i} G(p_i, t_i) \big\vert_{p_i = p_i^*(t_i,\cdot)},\\
&= 1-\frac{1}{\alpha_i}\mathbb{E}\left[H\left[ t_i - r_i(p_i^*(t_i,\cdot) ) \right] \right].
\end{aligned}
\end{equation}
Utilizing \eqref{eqn:t_cdf_R} on \eqref{eqn:t_subdiff_t}, we can show that a subgradient $g \in \partial_{t_i} G^* (t_i)$ of \eqref{eqn:t_subproblem} can be selected as
\begin{equation}
\label{eqn:t_F_R_alpha}
\hspace{-3pt} g \hspace{-1pt}=\hspace{-2pt} 
\begin{cases}
1, & \hspace{-4pt}\text{if } t_i < 0\\
1\hspace{-2pt}-\hspace{-2pt}\frac{1}{\alpha_i} C F_{h_i}\hspace{-2pt}\left( \sqrt{ \frac{\mu \alpha_i \sigma_i^2}{\lambda_i} } \right)\hspace{-1pt}, & \hspace{-4pt}\text{if } t_i = 0\\
1\hspace{-2pt}-\hspace{-2pt}\frac{1}{\alpha_i} F_{h_i}\hspace{-2pt}\left( \sqrt{ \frac{\mu \alpha_i \sigma_i^2}{\lambda_i} e^{t_i} } \right)\hspace{-1pt}, & \hspace{-4pt}\text{if } t_i > 0,
\end{cases}
\end{equation}
with $C \in [0,1]$ arbitrary. Notice that every such $g$ is decreasing and takes values in $[1, 1-\sfrac{1}{\alpha_i}]$, with a jump at $t_i = 0$. A subgradient satisfying $0 \in \partial_{t_i} G^*(t_i^*)$ can either occur when
\begin{equation}
\textstyle 1-\frac{1}{\alpha_i}F_{h_i}\left( \sqrt{ \frac{\mu \alpha_i \sigma_i^2}{\lambda_i} } \right) \geq 0 \Leftrightarrow \frac{\lambda_i}{\mu \alpha_i \sigma_i^2} \cdot \left( F_{h_i}^{-1}(\alpha_i) \right)^2 \geq 1,
\end{equation}
and $t_i^* = \log\left( \frac{\lambda_i}{\mu \alpha_i \sigma_i^2} \cdot \left( F_{h_i}^{-1}(\alpha_i) \right)^2 \right) \geq 0$ is a solution, or otherwise with the selection of the optimal $t_i^* = 0$, which concludes the proof.
\end{IEEEproof}
%\vspace{2pt}
For most standard fading distributions, e.g., Rayleigh, Weibull, Nakagami, Rician, Lognormal, etc., the particular solution of $\vect{t}$ uniquely exists. Further, with some tractable and analytically invertible distributions, e.g., Rayleigh and Weibull, the inverse of cdfs have tractable expressions which can be promptly leveraged to obtain simplest closed-form solutions for $\boldsymbol{t}^*$.

%For fading models with tractable and analytically invertible distributions, e.g., Rayleigh and Weibull distributions, pseudo-inverses correspond to particular expressions which can be promptly leveraged to obtain rigorous closed-form solutions for V@R. For most standard fading model Depending on the availability in the channel, we might simply recover the optimal $t_i$ either from the distribution of instantaneous rates or fading coefficients.
The last maximizing subproblem relates with the risk-ergodic rate $\vect{x}$. Nonetheless, it inherently depends on the concave objective function $f_0$, and dual variable $\vect{\lambda}$ --$\vect{\lambda}^{(n)}$ in a recursive fashion, $n \geq 0$--, such that
\begin{equation}
\label{eqn:x_closed_form}
\boxed{
\vect{x}^*(\vect{\lambda}) \in \arg \max_{\vect{x} \in \mathcal{X}}\ \left\{ f_0(\vect{x}) - \vect{\lambda}^T \vect{x} \right\}. }
\end{equation}
Again, we assume that such a solution as a function of $\vect{\lambda}$ exists, and $f_0$ is tractable, e.g., in closed-form, and readily available. Common objective functions inducing standard derivation and variable elimination, e.g., sumrate and proportional fairness utilities, are investigated later on.

\begin{algorithm}[tbp]
\centering
\begin{algorithmic}
\STATE Choose initial values $\boldsymbol{t}^{(0)}, \boldsymbol{p}^{(0)}, \boldsymbol{x}^{(0)}, \vect{\Lambda}^{(0)}$.
\FOR{$n = 1$ \textbf{to} Process End}
\STATE Observe $\boldsymbol{h}^{(n)}$.
\STATE \textit{\# Primal Variables}
%\STATE Calculate Branch Value by \eqref{eq:branch_cond}.
\STATE $\boldsymbol{\to}$ Set $t_i^*( \cdot )$ using \eqref{eqn:t_closed_form}, for all $i$.
\STATE $\boldsymbol{\to}$ Set $p_i^*\big( h_i^{(n)},\cdot \big)$ using \eqref{eqn:p_closed_form}, for all $i$.
\STATE $\boldsymbol{\to}$ Set $\boldsymbol{x}^*(\vect{\Lambda}^{(n-1)})$ using \eqref{eqn:x_closed_form}.
\STATE \textit{\# Dual Variables}
\STATE $\boldsymbol{\to}$ Update $\vect{\Lambda}^{(n)}$ using \eqref{eqn:dual_update} and \eqref{eqn:dual_subgradient}.
\ENDFOR
\end{algorithmic}
\caption{Dual Tail Waterfilling (DTW)}
\label{alg:dual_tail_waterfilling}
\end{algorithm}

\subsection{Dual Descent}

We are now left with the updates of the remaining dual variables, as all primal variables are explicitly expressed in closed-from as functions of dual variables. We might then restate the dual problem \eqref{eqn:dual_prob_explicit} with the optimal primal variables in place, as
\begin{equation}
\label{eqn:dual_descent_prob_formulation}
\begin{multlined}
D^* = \inf_{\vect{\Lambda} \succeq \vect{0}}\ f_0(\vect{x}^*)
+ \mu \left( P_0 - \Vert \mathbb{E}\left[ \vect{p}^*(\vect{h}) \right] \Vert_1 \right)\\
+\vect{\lambda}^T \left( \vect{t}^* - \frac{1}{\vect{\alpha}} \odot \mathbb{E}\left[ (\vect{t}^* - \vect{r}(\vect{p}^*(\vect{h}), \vect{h}) )_+ \right] - \vect{x}^* \right).
\end{multlined}
\end{equation}
Note that the dual function $D$ is convex with respect to $\vect{\Lambda} = (\vect{\lambda}, \mu)$. We then utilize the corresponding constraint gaps, in an analogous fashion to \cite{Ribeiro2012}, and formulate stochastic subgradient descent updates for dual variables $(\vect{\lambda},\mu)$, i.e.,
\begin{equation}
\label{eqn:dual_update}
\boxed{
\vect{\Lambda}^{(n)} = \left( \vect{\Lambda}^{(n-1)} - \varepsilon_{\vect{\Lambda}} g_{\vect{\Lambda}} \big(\vect{\Lambda}^{(n-1)} \big) \right)_+, }
\end{equation}
starting with $\vect{\Lambda}^{(0)}$, and where the stochastic subgradient vector $g_{\vect{\Lambda}}\big( \vect{\Lambda}^{(n-1)} \big) = \big[ g_{\vect{\lambda}}\big( \vect{\Lambda}^{(n-1)} \big) \, g_{\mu}\big( \vect{\Lambda}^{(n-1)} \big) \big]^T$ is expressed as
\begin{equation}
\label{eqn:dual_subgradient}
\begin{aligned}
g_{\vect{\vect{\lambda}}}\big( \vect{\Lambda}^{(n-1)} \big) &= \vect{t}^*\big( \vect{\Lambda}^{(n-1)} \big) - \frac{1}{\vect{\alpha}} \odot \Big( \vect{t}^*\big( \vect{\Lambda}^{(n-1)} \big)\\
&- \vect{r}\big( \vect{p}^*\big( \vect{h}^{(n)}, \vect{\Lambda}^{(n-1)} \big), \vect{h}^{(n)} \big) \Big)_+ - \vect{x}^*\big(\vect{\Lambda}^{(n-1)}\big),\\
g_{\mu} \big( \vect{\Lambda}^{(n-1)} \big) &= P_0 - \big\Vert \vect{p}^*\big( \vect{h}^{(n)}, \vect{\Lambda}^{(n-1)} \big) \big\Vert_1,
\end{aligned}
\end{equation}
with $\varepsilon_{\vect{\Lambda}}$ being a stepsize. Notice that \eqref{eqn:dual_subgradient} is a stochastic subgradient of the objective in \eqref{eqn:dual_descent_prob_formulation} from \cite[Theorem 7.52]{Shapiro2014}. Proposedly called \textit{dual tail waterfilling}, the complete characterization of the proposed dual descent scheme along with the parameterization of the primal variables, is presented in Algorithm~\ref{alg:dual_tail_waterfilling}.

\subsection{Common Utilities \& Fading Distributions}
\label{subsection:utilities}

We now examine several popular utilities and common fading distributions which are regularly practiced and investigated in applications.

\subsubsection{Sumrate Utility}

In case when $f_0(\vect{x}) = \vect{w}^T \vect{x},\ \vect{x} \in \mathbb{R}^{N_U}$ for an arbitrary weight vector $\vect{w} \in \mathbb{R}^{N_U}, \vect{w} \succ \vect{0}$, the subproblem with respect to risk-ergodic rate vector $\vect{x}$ becomes
\begin{equation}
\sup_{\vect{x} \in \mathcal{X}} \left\{(\vect{w} - \vect{\lambda})^T \vect{x} \right\},
\end{equation}
which is unbounded for any selection of $\vect{\lambda}$ and $\vect{w}$, except for the optimal selection of $\vect{\lambda}^* = \vect{w}$. This case inherently eliminates the steps for $\vect{x}$ and $\vect{\lambda}$.

\subsubsection{Proportional Fairness Utility}

In case when $f_0(\vect{x}) = \sum_{i=1}^{N_U} \log\left( x_i \right), \vect{x} \in \mathbb{R}^{N_U}$, the subproblem with respect to risk-ergodic rate vector $\vect{x}$ becomes
\begin{equation}
\sup_{\vect{x} \in \mathcal{X}} \left\{ \sum_{i=1}^{N_U} \log(x_i) - \lambda_i x_i \right\},
\end{equation}
which has a particular solution $\vect{x}^* = \frac{1}{\vect{\lambda}}$, emphasizing that the division by a vector stands for elementwise division.

For several popular fading models which enjoy favorable structure, i.e., analytical tractability and invertibility, the optimal $\vect{t}^*$ can be obtained purely in closed-form.

\setcounter{subsubsection}{0}

\subsubsection{Weibull Fading}

In case the channel follows a Weibull fading model, i.e., the cdfs of the $h_i$'s are
\begin{equation}
F_{h_i}(h;\nu_i,\kappa_i) = 1 - e^{-(\sfrac{h}{\nu_i})^{\kappa_i}}, \quad h \in [0,\infty),
\end{equation}
where $\nu_i$ is the scale parameter, and $\kappa_i$ is the shape parameter of the distribution. From \eqref{eqn:t_F_R_alpha}, we promptly arrive at
\begin{equation}
\boxed{
t_i^*(\lambda_i, \mu) = \left( \log\left( -\frac{2}{\kappa_i} \nu_i^2 \frac{\lambda_i}{\mu \alpha_i \sigma_i^2} \log\left( 1-\alpha_i \right) \right) \right)_+. }
\end{equation}

\subsubsection{Rayleigh Fading}

In this case the distribution functions of channel fading are described as
\begin{equation}
F_{h_i}(h; \rho_i) = 1 - e^{-\sfrac{h^2}{(2\rho_i^2)}}, \quad h \in [0,\infty),
\end{equation}
where $\rho_i$ is the scale parameter of the distribution. In fact, Rayleigh distribution is a particular case of Weibull distribution. From \eqref{eqn:t_F_R_alpha}, it then trivially follows that %for each $i \in \{0,\dots,N_U\}$ we have
\begin{equation}
\boxed{
t_i^*(\lambda_i, \mu) = \left(\log\left( -2\rho^2 \frac{\lambda_i}{\mu \alpha_i \sigma_i^2} \log\left( 1-\alpha_i \right) \right) \right)_+. }
\end{equation}

\section{Performance Evaluation}

We now confirm the effectiveness of the proposed dual tail waterfilling algorithm, presented in Algorithm~\ref{alg:dual_tail_waterfilling}. For the numerical simulations, we investigate a $3$-terminal point-to-point communication network consisting of independent --with no cross-interference-- links with distinct noise variance levels, operating under Rayleigh fading. The proposed dual tail waterfilling (DTW) algorithm is then applied with the utility functions presented in Section~\ref{subsection:utilities}, namely the sumrate and proportional fairness utilities.

\begin{table}[t]
    \centering
    \caption{Simulation parameters for $3$-terminal network}
    \begin{tabular}{cc||cc}
    \toprule
    \multicolumn{2}{c}{Sumrate} & \multicolumn{2}{c}{Proportional Fairness}\\
    \midrule
    $\vect{w}$ & $\begin{pmatrix} \sfrac{1}{3} & \sfrac{1}{3} & \sfrac{1}{3} \end{pmatrix}^T$ & &\\
    $\vect{\sigma}^2$ & $\begin{pmatrix} 1.0 & 2.0 & 3.0 \end{pmatrix}^T$ & $\vect{\sigma}^2$ & $\begin{pmatrix} 1.0 & 2.0 & 1.5 \end{pmatrix}^T$\\
    \midrule
    \multicolumn{2}{c}{$\vect{\rho}$} & \multicolumn{2}{c}{$\begin{pmatrix} 1 & 1 & 1\end{pmatrix}^T$}\\
    \multicolumn{2}{c}{$P_0, \varepsilon_{\vect{\Lambda}}, \varepsilon_{\vect{t}}$} & \multicolumn{2}{c}{$\begin{pmatrix}
    15 & 10^{-6} & 10^{-4}
    \end{pmatrix}^T$}\\
    \bottomrule
    \end{tabular}
    \label{tab:sim_parameters}
    \vspace{-4pt}
\end{table}

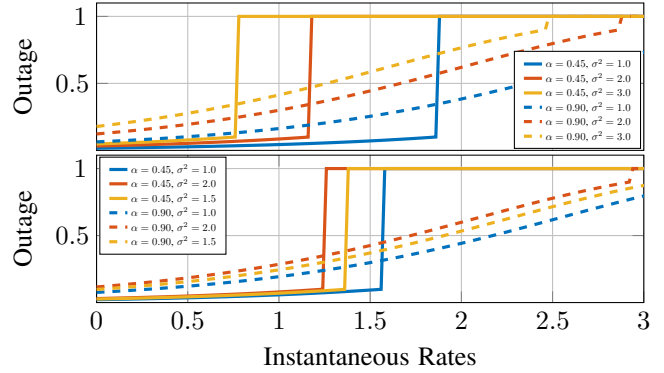
\begin{figure}[h]
    \centering
    \begin{subfigure}[ht]{\linewidth}
    \begin{tikzpicture}[trim axis right,baseline]
    \begin{axis}[
    width=\linewidth,
    height=.4\linewidth,
    ylabel={Outage}, %xlabel={Instantaneous Rates},
    xmin=0, xmax=3, xmajorticks=false,
    ymin=0, ymax=1.1, ytick = {0.5, 1},
    legend image post style={scale=0.4},
    legend style={at={(0.995,0.01)},anchor= south east,
    nodes={scale=0.4, transform shape}}, grid]
    \addplot[matblue, very thick] table[x=rate,y=cdf1] {Data/a010_sumrate_outages.txt};
    \addplot[matred, very thick] table[x=rate,y=cdf2] {Data/a010_sumrate_outages.txt};
    \addplot[matorange, very thick] table[x=rate,y=cdf3] {Data/a010_sumrate_outages.txt};
    \addplot[matblue, very thick, dashed] table[x=rate,y=cdf1] {Data/a090_sumrate_outages.txt};
    \addplot[matred, very thick, dashed] table[x=rate,y=cdf2] {Data/a090_sumrate_outages.txt};
    \addplot[matorange, very thick, dashed] table[x=rate,y=cdf3] {Data/a090_sumrate_outages.txt};
    \legend{$\alpha=0.45 \text{, } \sigma^2 = 1.0$, $\alpha=0.45 \text{, } \sigma^2 = 2.0$, $\alpha=0.45 \text{, } \sigma^2 = 3.0$, $\alpha=0.90 \text{, } \sigma^2 = 1.0$, $\alpha=0.90 \text{, } \sigma^2 = 2.0$, $\alpha=0.90 \text{, } \sigma^2 = 3.0$}
    \end{axis}
    \end{tikzpicture}
    \end{subfigure}
    \\
    \begin{subfigure}[ht]{\linewidth}
    \begin{tikzpicture}[trim axis right,baseline]
    \begin{axis}[
    width=\linewidth,
    height=.4\linewidth,
    ylabel={Outage}, xlabel={Instantaneous Rates},
    xmin=0, xmax=3, xmajorticks=true,
    ymin=0, ymax=1.1, ytick = {0.5, 1},
    legend image post style={scale=0.4},
    legend style={at={(0.005,0.99)},anchor=north west,
    nodes={scale=0.4, transform shape}}, grid]
    \addplot[matblue, very thick] table[x=rate,y=cdf1] {Data/a010_propfairness_outages.txt};
    \addplot[matred, very thick] table[x=rate,y=cdf2] {Data/a010_propfairness_outages.txt};
    \addplot[matorange, very thick] table[x=rate,y=cdf3] {Data/a010_propfairness_outages.txt};
    \addplot[matblue, very thick, dashed] table[x=rate,y=cdf1] {Data/a090_propfairness_outages.txt};
    \addplot[matred, very thick, dashed] table[x=rate,y=cdf2] {Data/a090_propfairness_outages.txt};
    \addplot[matorange, very thick, dashed] table[x=rate,y=cdf3] {Data/a090_propfairness_outages.txt};
    \legend{$\alpha=0.45 \text{, } \sigma^2 = 1.0$, $\alpha=0.45 \text{, } \sigma^2 = 2.0$, $\alpha=0.45 \text{, } \sigma^2 = 1.5$, $\alpha=0.90 \text{, } \sigma^2 = 1.0$, $\alpha=0.90 \text{, } \sigma^2 = 2.0$, $\alpha=0.90 \text{, } \sigma^2 = 1.5$}
    \end{axis}
    \end{tikzpicture}
    \end{subfigure}
    \caption{Outage probabilities for a $3$-terminal network with sumrate utility (top) and proportional fairness utility (bottom).}
    \vspace{2.5pt}
    \label{fig:outage}
\end{figure}

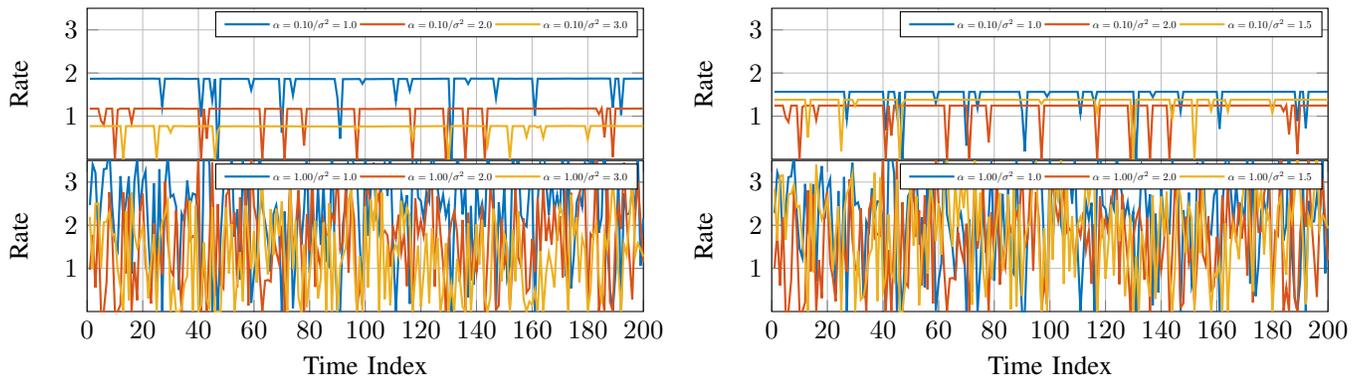
\begin{figure*}[ht]
\centering
    \begin{subfigure}[ht]{.495\linewidth}
    \begin{tikzpicture}[trim axis right,baseline]
    \begin{axis}[
    width=\linewidth,
    height=.4\linewidth,
    ytick={1,2,3,4},
    ylabel={Rate},
    xmin=0, xmax=200, ymin = 0, ymax=3.5, xmajorticks=false,
    legend style={at={(0.99,0.98)},anchor=north east,
    nodes={scale=0.4, transform shape}, legend columns = 3},
    grid]
    \addplot[matblue, thick, mark=no] table[x=index,y=rate1] {Data/a010_sumrate_rates.txt};
    \addplot[matred, thick, mark=no] table[x=index,y=rate2] {Data/a010_sumrate_rates.txt};
    \addplot[matorange, thick, mark=no] table[x=index,y=rate3] {Data/a010_sumrate_rates.txt};
    \legend{$\alpha = 0.10 / \sigma^2 = 1.0$, $\alpha = 0.10 / \sigma^2 = 2.0$, $\alpha = 0.10 / \sigma^2 = 3.0$}
    \end{axis}
    \end{tikzpicture}
    \end{subfigure}
    \begin{subfigure}[ht]{.495\linewidth}
    \begin{tikzpicture}[trim axis right,baseline]
    \begin{axis}[
    width=\linewidth,
    height=.4\linewidth,
    ytick={1,2,3,4},
    ylabel={Rate},
    xmin=0, xmax=200, ymin=0, ymax=3.5, xmajorticks=false,
    legend style={at={(0.99,0.98)},anchor=north east,
    nodes={scale=0.4, transform shape}, legend columns = 3}, grid]
    \addplot[matblue, thick, mark=no] table[x=index,y=rate1] {Data/a010_propfairness_rates.txt};
    \addplot[matred, thick, mark=no] table[x=index,y=rate2] {Data/a010_propfairness_rates.txt};
    \addplot[matorange, thick, mark=no] table[x=index,y=rate3] {Data/a010_propfairness_rates.txt};
    \legend{$\alpha = 0.10 / \sigma^2 = 1.0$, $\alpha = 0.10 / \sigma^2 = 2.0$, $\alpha = 0.10 / \sigma^2 = 1.5$}
    \end{axis}
    \end{tikzpicture}
    \end{subfigure}
    \begin{subfigure}[ht]{.495\linewidth}
    \begin{tikzpicture}[trim axis right,baseline]
    \begin{axis}[
    width=\linewidth,
    height=.4\linewidth,
    ylabel={Rate}, xlabel={Time Index}, 
    ytick={1,2,3,4},
    xmin=0, xmax=200, ymin=0, ymax=3.5, xmajorticks=true,
    legend style={at={(0.99,0.98)},anchor=north east,
    nodes={scale=0.4, transform shape}, legend columns = 3},
    grid]
    \addplot[matblue, thick, mark=no] table[x=index,y=rate1] {Data/a100_sumrate_rates.txt};
    \addplot[matred, thick, mark=no] table[x=index,y=rate2] {Data/a100_sumrate_rates.txt};
    \addplot[matorange, thick, mark=no] table[x=index,y=rate3] {Data/a100_sumrate_rates.txt};
    \legend{$\alpha = 1.00 / \sigma^2 = 1.0$, $\alpha = 1.00 / \sigma^2 = 2.0$, $\alpha = 1.00 / \sigma^2 = 3.0$}
    \end{axis}
    \end{tikzpicture}
    \end{subfigure}
    \begin{subfigure}[ht]{.495\linewidth}
    \begin{tikzpicture}[trim axis right,baseline]
    \begin{axis}[
    width=\linewidth,
    height=.4\linewidth,
    xlabel={Time Index}, ylabel={Rate},
    xmin=0, xmax=200, ymin=0, ymax=3.5,
    ytick={1,2,3,4},
    legend style={at={(0.99,0.98)},anchor=north east,
    nodes={scale=0.4, transform shape}, legend columns = 3}, grid]
    \addplot[matblue, thick, mark=no] table[x=index,y=rate1] {Data/a100_propfairness_rates.txt};
    \addplot[matred, thick, mark=no] table[x=index,y=rate2] {Data/a100_propfairness_rates.txt};
    \addplot[matorange, thick, mark=no] table[x=index,y=rate3] {Data/a100_propfairness_rates.txt};
    \legend{$\alpha = 1.00 / \sigma^2 = 1.0$, $\alpha = 1.00 / \sigma^2 = 2.0$, $\alpha = 1.00 / \sigma^2 = 1.5$}
    \end{axis}
    \end{tikzpicture}
    \end{subfigure}
    \caption{Achieved rates for the $3$-terminal network with sumrate utility (left), and proportional fairness utility (left). Top: risk-aware. Bottom: risk-neutral.}
    \label{fig:rates}
\end{figure*}

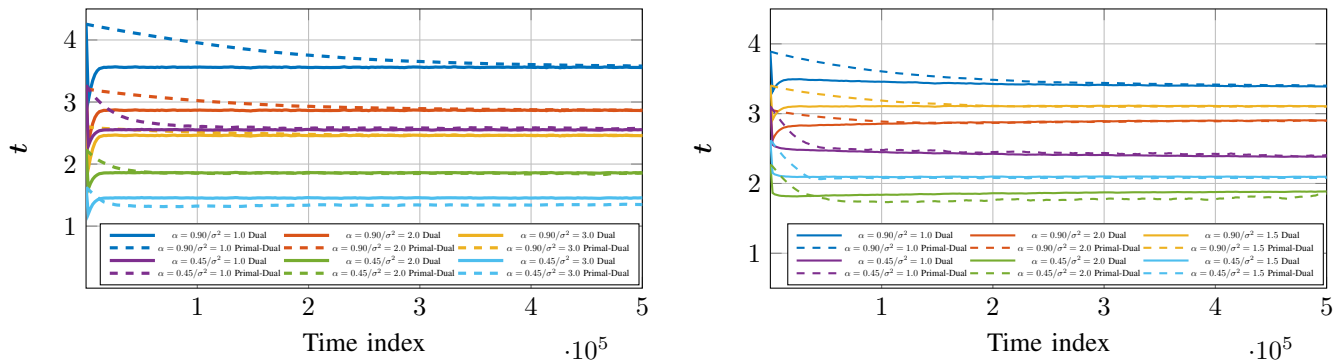
\begin{figure*}[h!]
\centering
    \begin{subfigure}[ht]{.495\linewidth}
    \begin{tikzpicture}[trim axis right,baseline]
    \begin{axis}[
    width=\linewidth,
    height=.59\linewidth,
    ytick={1,2,3,4},
    xtick={100000, 200000, 300000, 400000, 500000},
    ylabel={$\vect{t}$},
    xlabel={Time index},
    xmin=0, xmax=500000, ymin=0, ymax=4.5, xmajorticks=true,
    legend style={at={(0.995,0.01)},anchor=south east,
    nodes={scale=0.35, transform shape}, legend columns = 3},
    grid]
    \addplot[matblue, very thick, mark=no] table[x=index,y=t1] {Data/a090_sumrate_vars.txt};
    \addplot[matred, very thick, mark=no] table[x=index,y=t2] {Data/a090_sumrate_vars.txt};
    \addplot[matorange, very thick, mark=no] table[x=index,y=t3] {Data/a090_sumrate_vars.txt};
    \addplot[matblue, very thick, mark=no, dashed] table[x=index,y=t_pd1] {Data/a090_sumrate_vars.txt};
    \addplot[matred, very thick, mark=no, dashed] table[x=index,y=t_pd2] {Data/a090_sumrate_vars.txt};
    \addplot[matorange, very thick, mark=no, dashed] table[x=index,y=t_pd3] {Data/a090_sumrate_vars.txt};
    \addplot[matviolet, very thick, mark=no] table[x=index,y=t1] {Data/a045_sumrate_vars.txt};
    \addplot[matgreen, very thick, mark=no] table[x=index,y=t2] {Data/a045_sumrate_vars.txt};
    \addplot[matskyblue, very thick, mark=no] table[x=index,y=t3] {Data/a045_sumrate_vars.txt};
    \addplot[matviolet, very thick, mark=no, dashed] table[x=index,y=t_pd1] {Data/a045_sumrate_vars.txt};
    \addplot[matgreen, very thick, mark=no, dashed] table[x=index,y=t_pd2] {Data/a045_sumrate_vars.txt};
    \addplot[matskyblue, very thick, mark=no, dashed] table[x=index,y=t_pd3] {Data/a045_sumrate_vars.txt};
    \legend{$\alpha = 0.90 / \sigma^2 = 1.0$ Dual, $\alpha = 0.90 / \sigma^2 = 2.0$ Dual, $\alpha = 0.90 / \sigma^2 = 3.0$ Dual, $\alpha = 0.90 / \sigma^2 = 1.0$ Primal-Dual, $\alpha = 0.90 / \sigma^2 = 2.0$ Primal-Dual, $\alpha = 0.90 / \sigma^2 = 3.0$ Primal-Dual, $\alpha = 0.45 / \sigma^2 = 1.0$ Dual, $\alpha = 0.45 / \sigma^2 = 2.0$ Dual, $\alpha = 0.45 / \sigma^2 = 3.0$ Dual, $\alpha = 0.45 / \sigma^2 = 1.0$ Primal-Dual, $\alpha = 0.45 / \sigma^2 = 2.0$ Primal-Dual, $\alpha = 0.45 / \sigma^2 = 3.0$ Primal-Dual}
    \end{axis}
    \end{tikzpicture}
    \end{subfigure}
    \begin{subfigure}[ht]{.495\linewidth}
    \begin{tikzpicture}[trim axis right,baseline]
    \begin{axis}[
    width=\linewidth,
    height=.59\linewidth,
    ytick={1,2,3,4},
    xtick={100000, 200000, 300000, 400000, 500000},
    ylabel={$\vect{t}$},
    xlabel={Time index},
    xmin=0, xmax=500000, ymin=.5, ymax=4.5, xmajorticks=true,
    legend style={at={(0.995,0.01)},anchor=south east,
    nodes={scale=0.35, transform shape}, legend columns = 3}, grid]
    \addplot[matblue, thick, mark=no] table[x=index,y=t1] {Data/a090_propfairness_vars.txt};
    \addplot[matred, thick, mark=no] table[x=index,y=t2] {Data/a090_propfairness_vars.txt};
    \addplot[matorange, thick, mark=no] table[x=index,y=t3] {Data/a090_propfairness_vars.txt};
    \addplot[matblue, thick, mark=no, dashed] table[x=index,y=t_pd1] {Data/a090_propfairness_vars.txt};
    \addplot[matred, thick, mark=no, dashed] table[x=index,y=t_pd2] {Data/a090_propfairness_vars.txt};
    \addplot[matorange, thick, mark=no, dashed] table[x=index,y=t_pd3] {Data/a090_propfairness_vars.txt};
    \addplot[matviolet, thick, mark=no] table[x=index,y=t1] {Data/a045_propfairness_vars.txt};
    \addplot[matgreen, thick, mark=no] table[x=index,y=t2] {Data/a045_propfairness_vars.txt};
    \addplot[matskyblue, thick, mark=no] table[x=index,y=t3] {Data/a045_propfairness_vars.txt};
    \addplot[matviolet, thick, mark=no, dashed] table[x=index,y=t_pd1] {Data/a045_propfairness_vars.txt};
    \addplot[matgreen, thick, mark=no, dashed] table[x=index,y=t_pd2] {Data/a045_propfairness_vars.txt};
    \addplot[matskyblue, thick, mark=no, dashed] table[x=index,y=t_pd3] {Data/a045_propfairness_vars.txt};
    \legend{$\alpha = 0.90 / \sigma^2 = 1.0$ Dual, $\alpha = 0.90 / \sigma^2 = 2.0$ Dual, $\alpha = 0.90 / \sigma^2 = 1.5$ Dual, $\alpha = 0.90 / \sigma^2 = 1.0$ Primal-Dual, $\alpha = 0.90 / \sigma^2 = 2.0$ Primal-Dual, $\alpha = 0.90 / \sigma^2 = 1.5$ Primal-Dual, $\alpha = 0.45 / \sigma^2 = 1.0$ Dual, $\alpha = 0.45 / \sigma^2 = 2.0$ Dual, $\alpha = 0.45 / \sigma^2 = 1.5$ Dual, $\alpha = 0.45 / \sigma^2 = 1.0$ Primal-Dual, $\alpha = 0.45 / \sigma^2 = 2.0$ Primal-Dual, $\alpha = 0.45 / \sigma^2 = 1.5$ Primal-Dual}
    \end{axis}
    \end{tikzpicture}
    \end{subfigure}
    \caption{$\vect{t}$-iterates for the $3$-terminal network with sumrate utility (left), and proportional fairness utility (right).}
    \label{fig:optimal_vars}
    \vspace{-12pt}
\end{figure*}

%We might be able to recover the distribution of instantaneous rates in \eqref{eqn:t_cdf_R} by leveraging the optimal resource policy \eqref{eqn:p_closed_form}, assuming that the fading distribution is provided.
The outage probability, i.e., $\mathbb{P}_{\mathrm{out}}( r_o ) = \mathbb{P}\left\{ R \leq r_0 \right\}$, can naturally be taken as another instructive measure of robustness. %By the formulation of the cumulative distribution function \eqref{eqn:t_cdf_R}, 
The CV@R-optimal instantaneous rates exhibit a sharp statistical threshold, as shown in Fig.~\ref{fig:outage}, due to the risk-averse rate-constraining nature of the CV@R. For smaller values of $\vect{\alpha}$ --corresponding to stricter, more conservative risk-aware settings-- the outage probability is substantially lower at always-attainable rate levels. Conversely, larger $\vect{\alpha}$ values --corresponding to less risk-aware settings-- induce much higher variability in optimal instantaneous rates, see Fig.~\ref{fig:rates} (bottom). The confidence level $\vect{\alpha}$ concurrently regulates the distribution of rates and the instantaneous rate boundary $\vect{t}^*$.

To further elaborate on the efficacy of the proposed dual tail waterfilling (DTW) algorithm, we compare it with the primal-dual tail waterfilling (PDTW) algorithm developed in \cite{yaylali2023cvar}, in terms of convergence to the optimal $\vect{t}^*$. %Numerical simulations shown in Fig.~\ref{fig:optimal_vars} demonstrate that closed-form solution of optimal V@R converges at a faster rate than the variable $\vect{t}$ in the primal-dual scheme. %Clearly seen in the figures, the primal-dual scheme takes much longer to convergence in $\vect{t}$, and even when the stepsize $\varepsilon_{\vect{t}}$ is larger, the fluctuations might result in either promptly diverging from the optimal and/or incurring instability to the system. 
DTW uses additional statistical information (i.e., fading distributions) to obtain closed-form expressions for $\vect{t}^*$ relative to dual variables, which is observed to converge rapidly --see Fig.~\ref{fig:optimal_vars}--, (provided that fading distributions are known). On the other extent, PDTW leverages a purely data-driven scheme to learn globally optimal primal and dual variables. In both methods, the CV@R level $\vect{\alpha}$ constraints the attainable rates to lower $\vect{\alpha}$-quantiles, particularly upper bounded by $\vect{t}^*$. For a small $\vect{\alpha}$, the variable $\vect{t}$ drastically limits the achievable rates, and immensely increases their probability of eventually attaining the optimal $\vect{t}^*$. Since stochastic subgradient ascent for $\vect{t}$ depends on instantaneous rates/fading realizations --as it happens for PDTW \cite[Section IV-B]{yaylali2023cvar}--, such a data-driven approach will be susceptible to diverging for small values of $\alpha$ due to data starvation (increasing rarity of ``bad" fading events). This issue is not observed in the proposed DTW algorithm even for rather small values for $\vect{\alpha}$, as DTW leverages knowledge of the fading distributions through our explicit closed-form expressions for $\vect{t}^*$.

\section{Conclusion}
\vspace{-0pt}
We investigated a risk-aware formulation of a classical but fundamental stochastic resource allocation problem in point-to-point communication networks. Exploiting CV@R as a measure of risk, we proposed dual tail waterfilling (DTW), a purely dual version the primal-dual tail waterfilling (PDTW) algorithm recently proposed in \cite{yaylali2023cvar}. We developed closed-form solutions for all primal variables, and derived stochastic subgradient updates for dual variables. Detailed numerical simulations implemented over two typical utilities effectively corroborated the efficacy and characteristics of the proposed algorithm, as well as the precise and rapid global convergence in both primal and dual variables. 

\bibliographystyle{IEEEtran}
\bibliography{Files/references}

\end{document}